\documentclass[a4paper,11pt]{article}
\usepackage{pos}

\title{Status update: $\pi^0\to \gamma^\ast \gamma^\ast$ transition form factor on CLS ensembles}

\author*[a]{Jonna~Koponen}
\author[b]{Antoine G\'erardin}
\author[a,c]{Harvey~B.~Meyer}
\author[a]{Konstantin Ottnad}
\author[a]{Georg von Hippel}

\affiliation[a]{PRISMA$^+$ Cluster of Excellence \& Institut f\"{u}r Kernphysik,
  Johannes-Gutenberg-Universit\"{a}t  Mainz,\\  D-55099 Mainz, Germany}

\affiliation[b]{Aix-Marseille Universit\'e, Universit\'e de Toulon, CNRS, CPT,\\ Marseille, France}

\affiliation[c]{Helmholtz-Institut Mainz, Johannes-Gutenberg-Universit\"{a}t Mainz,\\
D-55099 Mainz, Germany}

\emailAdd{jkoponen@uni-mainz.de}

\abstract{
  In this report we present the status of the Mainz group's lattice QCD calculation of the pion
  transition form factor, which describes the interaction of an on-shell
  pion with two off-shell photons. This form factor is the main ingredient in the calculation
  of the pion-pole contribution to hadronic light-by-light scattering in the muon $g-2$.

  We use the $N_f = 2 + 1$ CLS gauge ensembles, and we update our previous work by including a
  physical pion mass ensemble (E250). We compute the transition form factor in a moving frame
  as well as in the pion rest frame in order to have access to a wider range of photon virtualities.
  In addition to the quark-line connected correlator we also compute the quark-line disconnected
  diagrams that contribute to the form factor.

  At the final stage of the analysis, the result on E250 will be combined with the previous work
  published in 2019 to extrapolate the form factor to the continuum and to physical quark masses.
}

\FullConference{The 40th International Symposium on Lattice Field Theory (Lattice 2023)\\
July 31st - August 4th, 2023\\
Fermi National Accelerator Laboratory\\}

\begin{document}
\maketitle

\section{Introduction and motivation}

The transition form factor (TFF) $\mathcal{F}_{\pi^0\gamma^\ast\gamma^\ast}$ describes the interaction of an
on-shell pion with two off-shell photons. It is the main ingredient in the calculation of the pion-pole
contribution to hadronic light-by-light scattering in the muon $g-2$. There is also a direct
relation between $\mathcal{F}_{\pi^0\gamma^\ast\gamma^\ast}(0,0)$ (the transition form factor with two real photons)
and the partial decay width $\Gamma (\pi^0\to \gamma\gamma)$
(see Eq.~\eqref{eq:TFFandPartialDecayWidth} in Section~\ref{sec:results}). In
the leading order of chiral perturbation theory ($\chi$PT),
\begin{equation}
  \Gamma(\pi^0\to \gamma\gamma) = \dfrac{m^3_{\pi^0}\alpha_e^2N_c^2}{576\pi^3 F^2_{\pi^0}},
\end{equation}
where $\alpha_e$ is the fine structure constant, $N_c$ is the number of colors and $F_{\pi^0}$ is pion decay
constant in the chiral limit.
There is a tension between the measured value and the theoretical predictions when NLO corrections are added,
which is illustrated very clearly for example in Fig. 72 in \cite{Accardi:2023chb}. Lattice calculations could
shed light to this issue, if 2\% precision can be achieved on the normalization of the TFF.

\section{Extraction of the TFF}

The calculation follows very closely the Mainz group's publication~\cite{Gerardin:2019vio} (see also
\cite{Gerardin:2016cqj}). The transition form factor is extracted from matrix elements
\begin{equation}
  M_{\mu\nu}(p,q_1)=\mathit{i}\!\int\! \mathrm{d}^4x \mathrm{e}^{\mathit{i}q_1\cdot x}\langle 0 |T\{J_\mu(x)J_\nu(0)\}|\pi^0(p)\rangle 
  = \epsilon_{\mu\nu\alpha\beta}q^\alpha_1q^\beta_2\mathcal{F}_{\pi^0\gamma^\ast\gamma^\ast}(q^2_1,q^2_2),
\end{equation}
where $J_\mu$ is the electromagnetic (EM) current. Here $q_1 = (\omega_1,\vec{q}_1)$ and
$q_2 = (E_\pi-\omega_1,\vec{p}-\vec{q}_1)$ are the four-momenta associated with the two
currents, and $p$ is the four-momentum of the pion, such that $p=q_1+q_2$.

\begin{figure}
\includegraphics[trim={3mm 2mm 5mm 2mm},clip,width=0.49\linewidth]{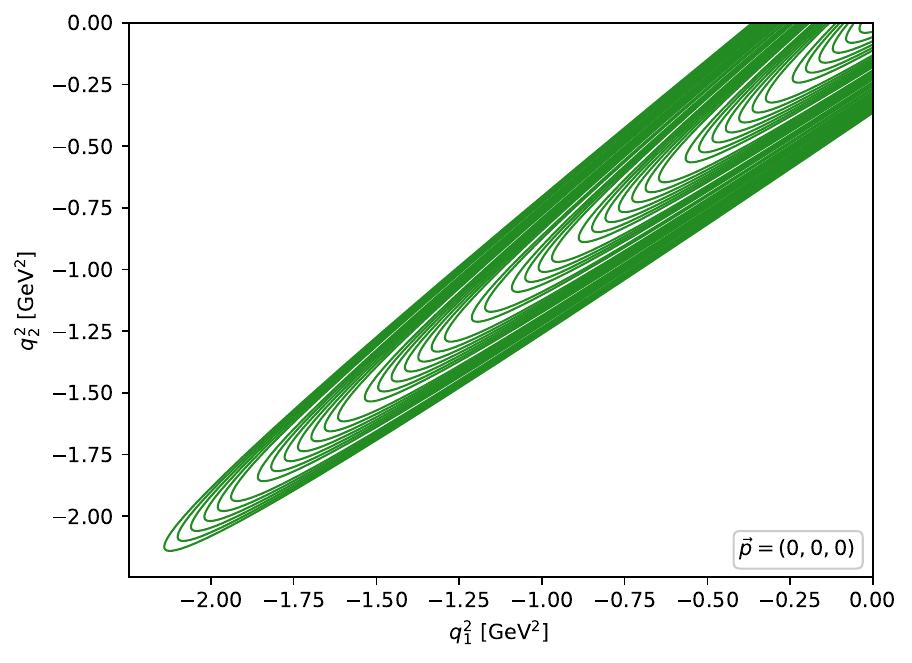}
\includegraphics[trim={1mm 2mm 5mm 2mm},clip,width=0.49\linewidth]{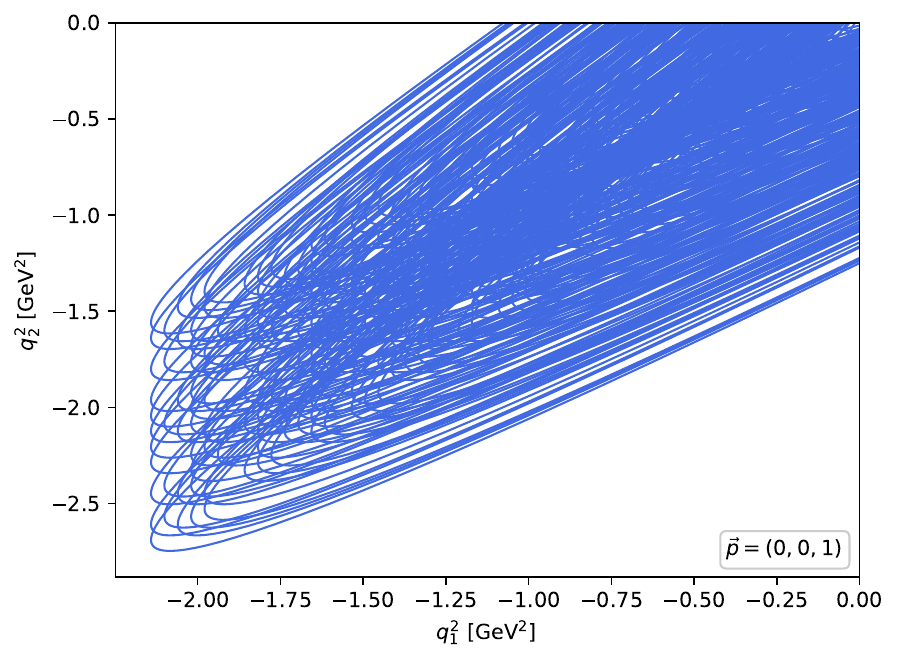}
\caption{Photon virtualities at the physical pion mass with $L\approx 6$~fm.}
\label{fig:photon_virtualities}
\end{figure}

To cover a wide range of photon virtualities, we use both the rest frame of the pion, $\vec{p}=(0,0,0)$, and a moving
frame $\vec{p}=(0,0,1)$ (in units of $2\pi/L$). The increase in the range of accessible virtualities is best illustrated
in the $(q_1^2,q_2^2)$-plane --- see Fig.~\ref{fig:photon_virtualities}. Each curve in the plot represents a fixed
value of $\vec{q}_1$ and $\vec{p}$, and $\omega_1$ is a free parameter (this tracks the curve from one end to another).

\begin{figure}[b]
\includegraphics[trim={0mm 0mm 0mm 0mm},clip,width=0.995\linewidth]{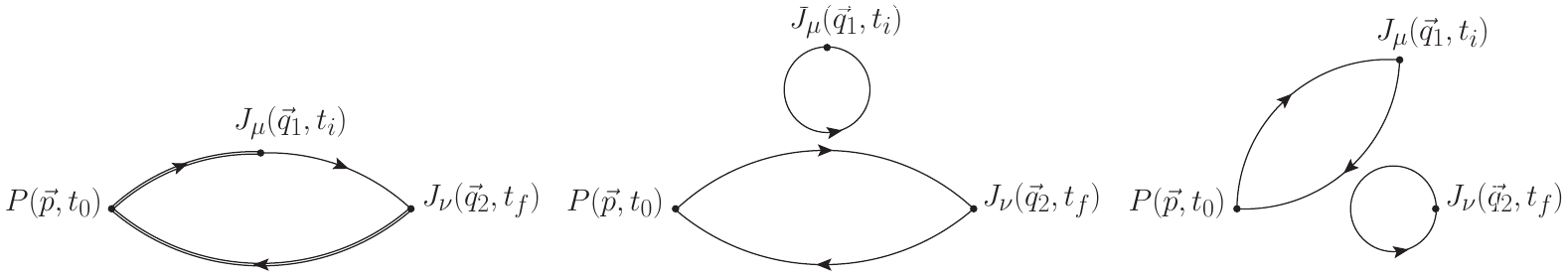}
\caption{Connected and disconnected diagrams.}
\label{fig:diagrams}  
\end{figure}  

The Euclidean matrix elements read
\begin{equation}
M_{\mu\nu} = (\mathit{i}^{n_0}) M^E_{\mu\nu},\;
M^E_{\mu\nu} = -\!\int^\infty_{-\infty}\!\mathrm{d}\tau \mathrm{e}^{\omega_1\tau}\!\int\! \mathrm{d}^3z \mathrm{e}^{-\mathit{i}\vec{q}_1\cdot \vec{x}}\langle 0|T\{J_\mu(\vec{x},\tau)J_\nu(\vec{0},0)\}|\pi^0(p)\rangle,
\end{equation}
where $n_0$ denotes the number of temporal indices.
The matrix elements can be obtained by integration over an Euclidean time dependent amplitude,
\begin{equation}
  M^E_{\mu\nu}(p,q_1)= \frac{2E_\pi}{Z_\pi}\!\int^\infty_{-\infty}\!\mathrm{d}\tau \mathrm{e}^{\omega_1\tau}\widetilde{A}_{\mu\nu}(\tau),
\end{equation}
where $\tau=t_i-t_f$ is the time separation between the two EM currents.
The amplitude $\widetilde{A}_{\mu\nu}(\tau)$ is connected to a 3-point correlator calculated on the lattice by
\begin{align}
  C^{(3)}_{\mu\nu}(\tau,t_\pi) &\equiv a^6\sum_{\vec{x},\vec{z}}\langle J_{\mu}(\vec{x},t_i)J_{\nu}(\vec{0},t_f)P^{\dag}(\vec{z},t_0)\rangle \mathrm{e}^{\mathit{i}\vec{p}\cdot\vec{z}}\mathrm{e}^{-\mathit{i}\vec{q}_1\cdot\vec{x}}\\
  \widetilde{A}_{\mu\nu}(\tau)&\equiv \lim_{t_\pi\to +\infty}\mathrm{e}^{E_\pi(t_f-t_0)}C^{(3)}_{\mu\nu}(\tau,t_\pi),\; (t_0 < t_f),
\end{align}
where $t_\pi$ is the time separation between the pion and the closest EM current.
In addition to the quark-line connected diagram, there are contributions from two quark-line disconnected diagrams that
have to be calculated. Both the connected and disconnected diagrams are depicted in Fig.~\ref{fig:diagrams}.

\subsection{Lattice ensembles}

We use the CLS $N_f=2+1$ ensembles with non-perturbatively $\mathcal{O}(a)$-improved Wilson fermions and
tree-level improved L\"uscher-Weisz gauge action. We have four lattice spacings and use multiple pion masses
to control the chiral extrapolation. All ensembles have fairly large volumes ($M_{\pi}L\ge 4$). More details
about the ensembles can be found in~\cite{Gerardin:2019vio} and references therein. Compared to the
publication~\cite{Gerardin:2019vio} we now add one ensemble (E250) at the physical pion mass with a lattice
spacing of $a\approx 0.064$~fm, size $96^3\times 192$ and $L\approx 6$~fm.

\subsection{Correlators}

\begin{figure}[b]
\includegraphics[trim={0mm 2mm 16mm 14mm},clip,width=0.49\textwidth]{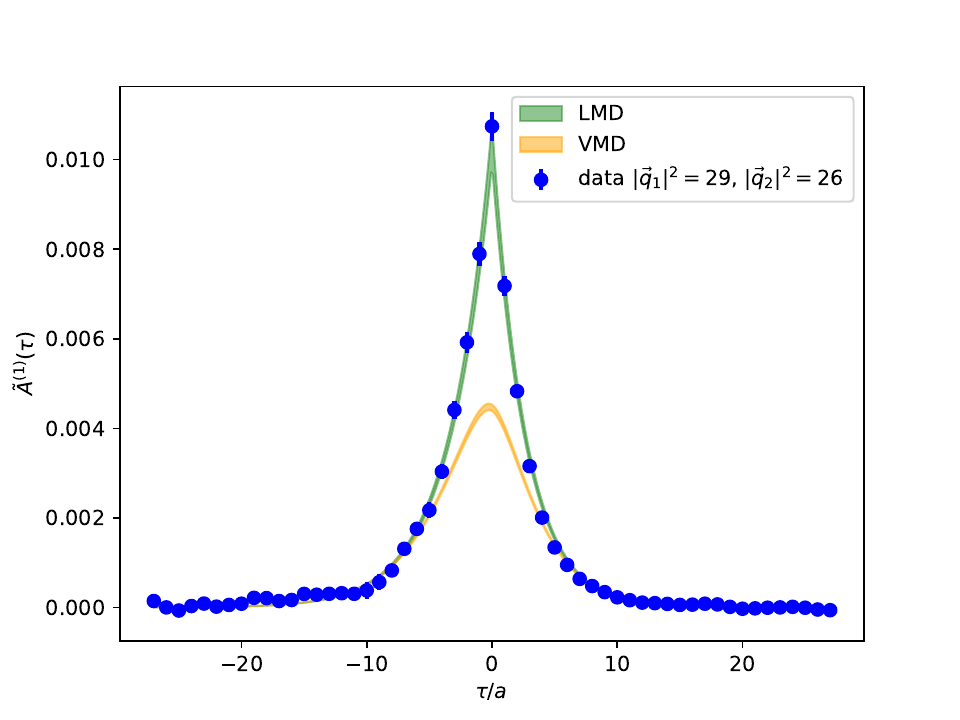}
\includegraphics[trim={0mm 2mm 16mm 14mm},clip,width=0.49\textwidth]{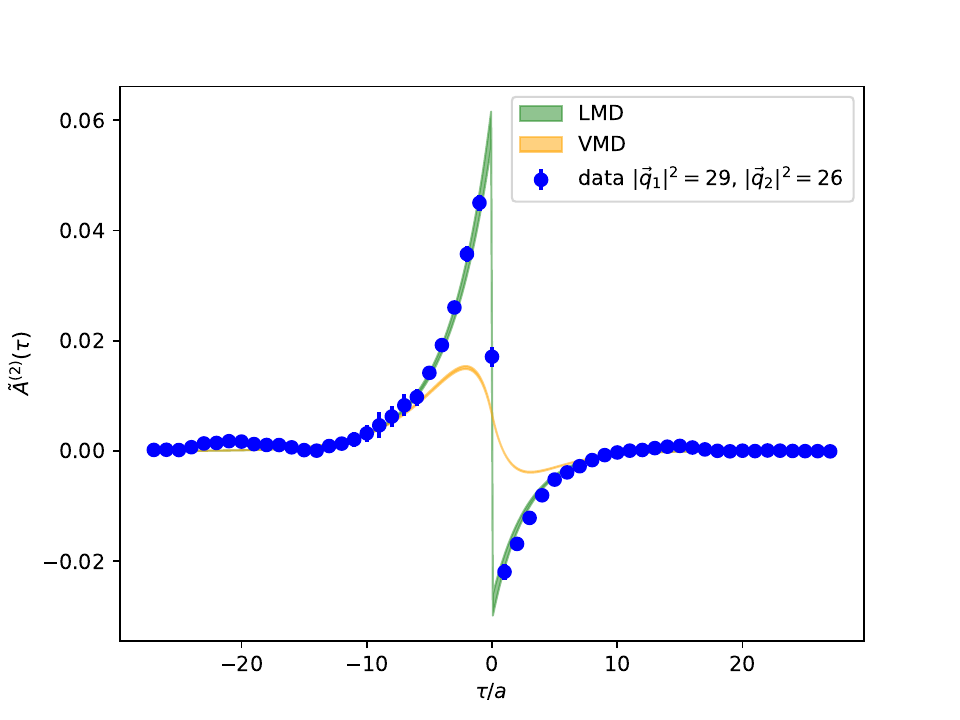}
\caption{Examples of the connected contribution to the scalar functions $\widetilde{A}^{(1)}(\tau)$
  and $\widetilde{A}^{(2)}(\tau)$. The momenta are given in units of $2\pi/L$.}
\label{fig:A1_and_A2}
\end{figure}

Recall that $\widetilde{A}_{\mu\nu}(\tau)$ is directly related to the 3-point correlators $C^{(3)}_{\mu\nu}$.
For convenience we define two scalar funtions $\widetilde{A}^{(1)}(\tau)$ and $\widetilde{A}^{(2)}(\tau)$:
\begin{align}
\widetilde{A}_{0k}(\tau) =& (\vec{q}_1\times \vec{p})_k\widetilde{A}^{(1)}(\tau), \nonumber \\
{\epsilon'}^k \widetilde{A}_{kl}(\tau)\epsilon^l =& -\mathit{i}(\vec{\epsilon}'\times \vec{\epsilon})
\cdot\left(\vec{q}_1E_\pi\widetilde{A}^{(1)}(\tau) + \vec{p}\frac{\mathrm{d}\widetilde{A}^{(1)}(\tau)}{\mathrm{d}\tau}\right).
\end{align}
In the moving frame we define for simplicity $\widetilde{A}_{12}(\tau) \equiv -\mathit{i}E_\pi p_z\widetilde{A}^{(2)}(\tau)$.
The two scalar functions have very distinct features as can be seen in Fig.~\ref{fig:A1_and_A2}: $\widetilde{A}^{(1)}$
peaks at $\tau=0$, whereas $\widetilde{A}^{(2)}$ changes sign. See section~\ref{sec:tail} for more details of the two fits,
VMD and LMD, that are used to model the tail contribution.

\subsection{Disconnected contribution}

\begin{figure}[t]
\includegraphics[trim={1mm 0mm 2mm 2mm},clip,width=0.49\textwidth]{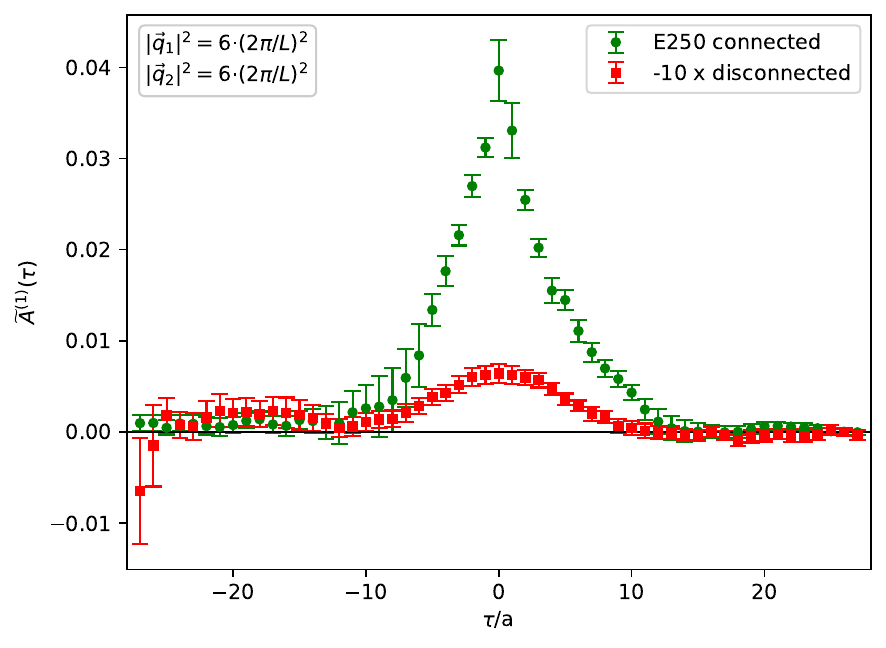}
\includegraphics[trim={1mm 0mm 2mm 2mm},clip,width=0.49\textwidth]{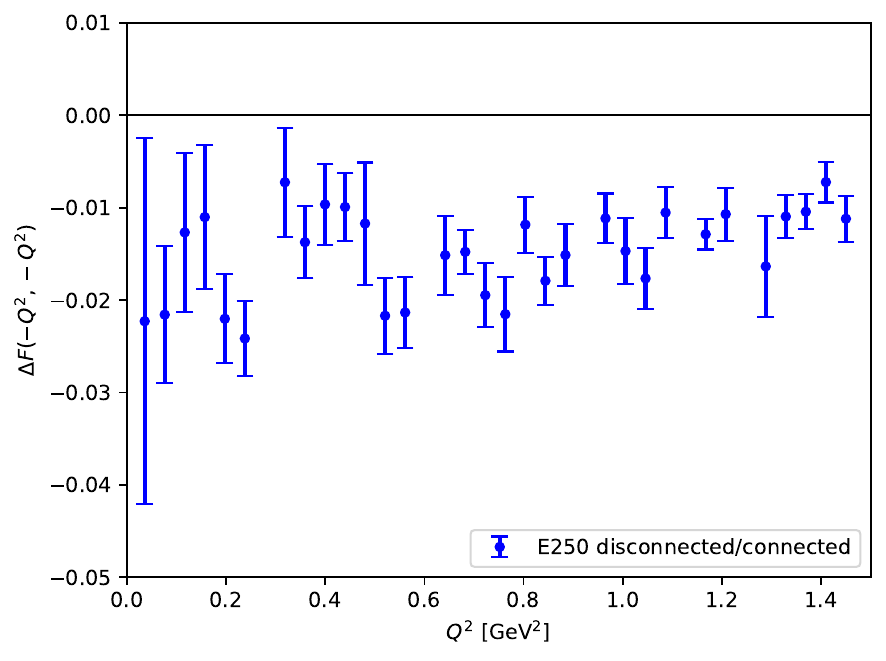}
\caption{Disconnected contribution.}
\label{fig:disconnected}
\end{figure}

In addition to the quark-line connected piece, we need two quark-line disconnected diagrams.
The quark loops are computed using stochastic all-to-all methods \cite{Giusti:2019kff, Jansen:2008wv, Stathopoulos:2013aci},
while the two-point functions are computed using point sources.
We find the disconnected contribution
\begin{equation}
  \Delta F(-Q^2_1,-Q^2_2)=\dfrac{\mathcal{F}^{\textrm{disc}}_{\pi^0\gamma^\ast \gamma^\ast}(-Q^2_1,-Q^2_2)}{\mathcal{F}^{\textrm{conn}}_{\pi^0\gamma^\ast \gamma^\ast}(-Q^2_1,-Q^2_2)}
\label{eq:disc_fraction}  
\end{equation}
is at the few-percent level. This is illustrated in Fig.~\ref{fig:disconnected}, where the plot on the left shows the connected
piece and the disconnected piece (multiplied by ten) for the scalar function $\widetilde{A}^{(1)}$, whereas the plot on the right
shows the ratio given in Eq.~\eqref{eq:disc_fraction}.

\subsection{Modeling the tail}
\label{sec:tail}

Recall that we extract the TFF by evaluating the integral
$\int^\infty_{-\infty}\!\!\mathrm{d}\tau \mathrm{e}^{\omega_1\tau}\widetilde{A}_{\mu\nu}(\tau).$ We need to
model $\widetilde{A}_{\mu\nu}(\tau)$ at large $|\tau|$ to get the tail contribution. We use two models:

\begin{itemize}
\item Lowest Meson Dominance (LMD)

\begin{equation}
    \widetilde{A}^{\textrm{LMD}}_{\mu\nu}(\tau)=\frac{Z_\pi}{4\pi E_\pi}\int_{-\infty}^{\infty}\mathrm{d}\widetilde{\omega}
\dfrac{\left(P_{\mu\nu}^E\widetilde{\omega}+Q^E_{\mu\nu}\right)\left(\alpha M_V^4+\beta(q_1^2+q_2^2)\right)}{\left(\widetilde{\omega}-\widetilde{\omega}_1^{(+)}\right)\left(\widetilde{\omega}-\widetilde{\omega}_1^{(-)}\right)\left(\widetilde{\omega}-\widetilde{\omega}_2^{(+)}\right)\left(\widetilde{\omega}-\widetilde{\omega}_2^{(-)}\right)}\mathrm{e}^{-\mathit{i}\widetilde{\omega}\tau}
\end{equation}
with
\begin{align*}
P_{\mu\nu}^E=&\mathit{i}\epsilon_{\mu\nu0i}p^i,& \widetilde{\omega}_1^{(\pm)}&=\pm\mathit{i}\sqrt{M^2_V+|\vec{q}_1|^2},\\
  Q^E_{\mu\nu}=&\epsilon_{\mu\nu i0}E_\pi q_1^i-\mathit{i}\epsilon_{\mu\nu ij}q^i_1p^j,& \widetilde{\omega}_2^{(\pm)}&=-\mathit{i}\left(E_\pi \mp\sqrt{M^2_V+|\vec{q}_2|^2}\right),
\end{align*}
and $Z_\pi$ the overlap factor from the pion 2-point function.
This gives an explicit expression for $\widetilde{A}^{\textrm{LMD}}_{\mu\nu}$, which we use to
fit our data using $\alpha$, $\beta$ and $M_V$ as fit parameters.

\item Vector Meson Dominance (VMD): Set $\beta=0$ in the LMD model
  
\end{itemize}

These models are used at $|\tau|/a > 20$ ($|\tau| > 1.3$~fm), to make sure the contribution
from the tail is small compared to the contribution we extract directly from our lattice data.
Both VMD and LMD model fits to the data are shown in Fig.~\ref{fig:A1_and_A2}. Especially the
latter model describes the data well even fairly close to $\tau = 0$, and we include data
$|\tau|/a \geq 8$ ($|\tau| \geq 0.5$~fm) in these fits.

\subsection{Parameterizing the form factor: $z$-expansion}

After obtaining the transition form factor at several virtualities
$(q_1^2,q_2^2)\equiv (-Q_1^2,-Q_2^2)$, we parameterize it using a conformal mapping
\begin{equation}
  z_{k} = \dfrac{\sqrt{t_{\textrm{cut}}+Q^2_k}-\sqrt{t_{\textrm{cut}}-t_0\vphantom{t_{\textrm{cut}}+Q^2_{k}}}}{\sqrt{t_{\textrm{cut}}+Q^2_{k}}+\sqrt{t_{\textrm{cut}}-t_0\vphantom{t_{\textrm{cut}}+Q^2_{k}}}},\textrm{ with } t_{\textrm{cut}}=4(m_\pi^\textrm{phys})^2,\textrm{ and } t_0=t_{\textrm{cut}}\left(1-\sqrt{1+\dfrac{Q^2_{\textrm{max}}}{t_{\textrm{cut}}}}\right).
\end{equation}  
The form factor is then written as an expansion in $z_1$ and $z_2$:
\begin{multline}  
    P(Q_1^2,Q_2^2)\mathcal{F}_{\pi^0 \gamma^\ast \gamma^\ast}(-Q^2_1,-Q^2_2)=\\
    \sum_{n,m=0}^{N}c_{nm}\left(z^n_1+(-1)^{N+n}\dfrac{n}{N+1}z_1^{N+1}\right)\left(z^m_2+(-1)^{N+m}\dfrac{m}{N+1}z_2^{N+1}\right), 
\end{multline}
where the coefficients $c_{nm}=c_{mn}$, the fit parameters, are symmetric. The polynomial
\begin{equation}
  P(Q_1^2,Q_2^2)=1+\dfrac{Q_1^2+Q_2^2}{M_V^2}
\end{equation}
implements the vector meson pole with $M_V=775$~MeV and ensures the correct asymptotic behaviour at
large virtualities.

\section{Results}
\label{sec:results}

\begin{figure}
\includegraphics[trim={0mm 0mm 0mm 0mm},clip,width=0.49\textwidth]{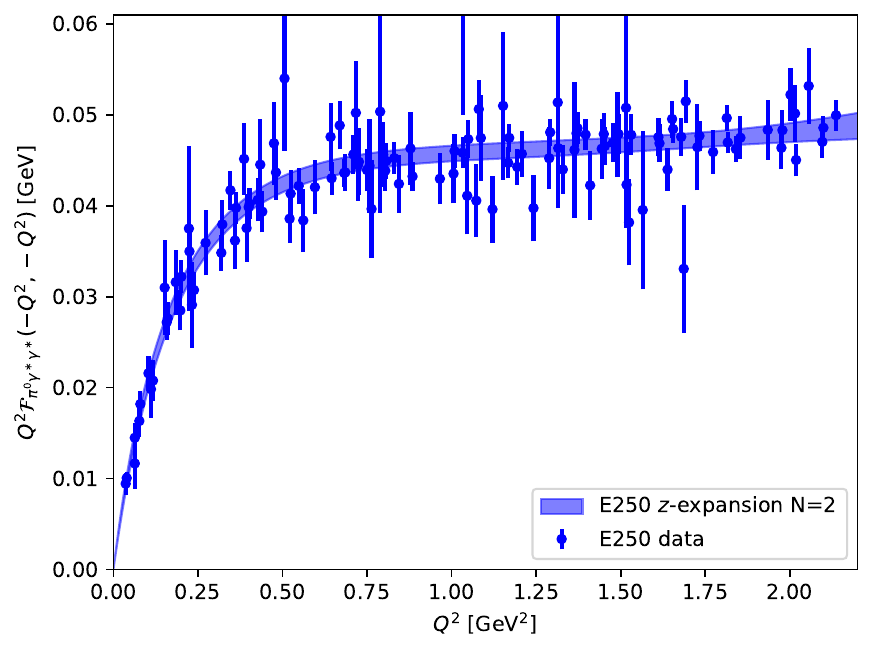} 
\includegraphics[trim={0mm 0mm 0mm 0mm},clip,width=0.49\textwidth]{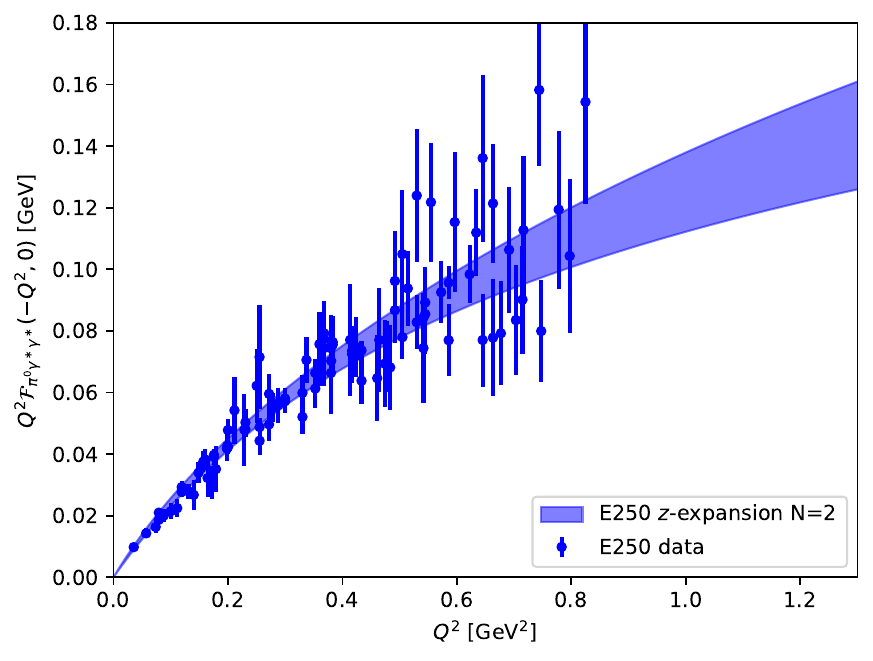} 
\caption{Transition form factor $Q^2\mathcal{F}_{\pi^0\gamma^\ast \gamma^\ast}$. On the left: double virtual
  $Q^2\mathcal{F}_{\pi^0 \gamma^\ast \gamma^\ast}(-Q^2,-Q^2)$; on the right: single virtual $Q^2\mathcal{F}_{\pi^0 \gamma^\ast \gamma^\ast}(-Q^2,0)$.}
\label{fig:TFF}  
\end{figure}

Preliminary results for the transition form factor on the physical pion mass ensemble are plotted in Fig.~\ref{fig:TFF}
along with the $N=2$ $z$-expansion. We plot our fit for two specific choices of $\omega_1$, that correspond to the
double virtual $Q^2_1=Q^2_2=Q^2$ and single virtual $Q^2_1=Q^2$, $Q^2_2=0$, cases. The $z$-expansion with $N=2$ clearly
describes the data well.

\begin{figure}
\includegraphics[trim={0mm 0mm 0mm 0mm},clip,width=0.49\textwidth]{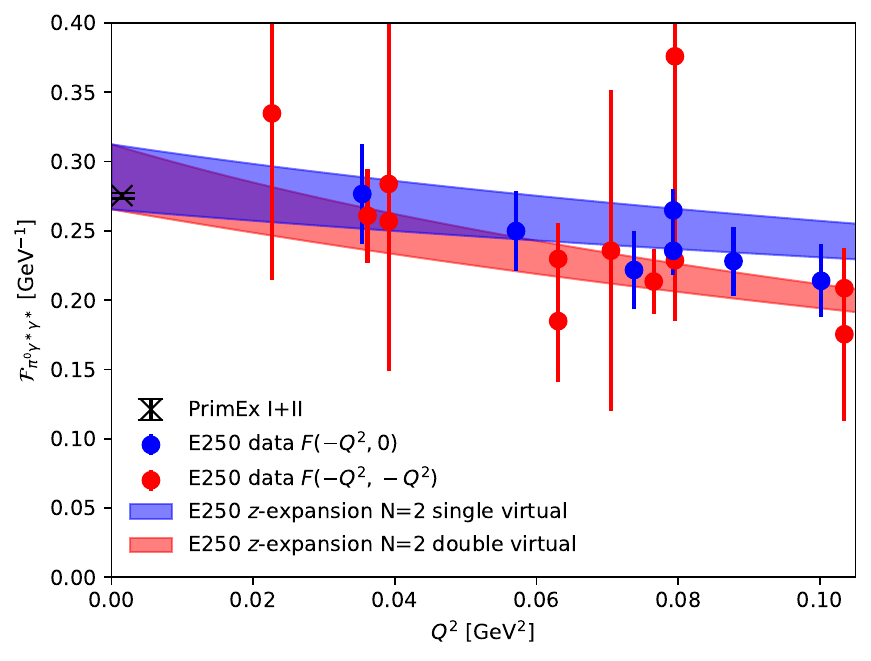}
\includegraphics[trim={-4mm -9.5mm 0.7mm 0mm},clip,width=0.49\textwidth]{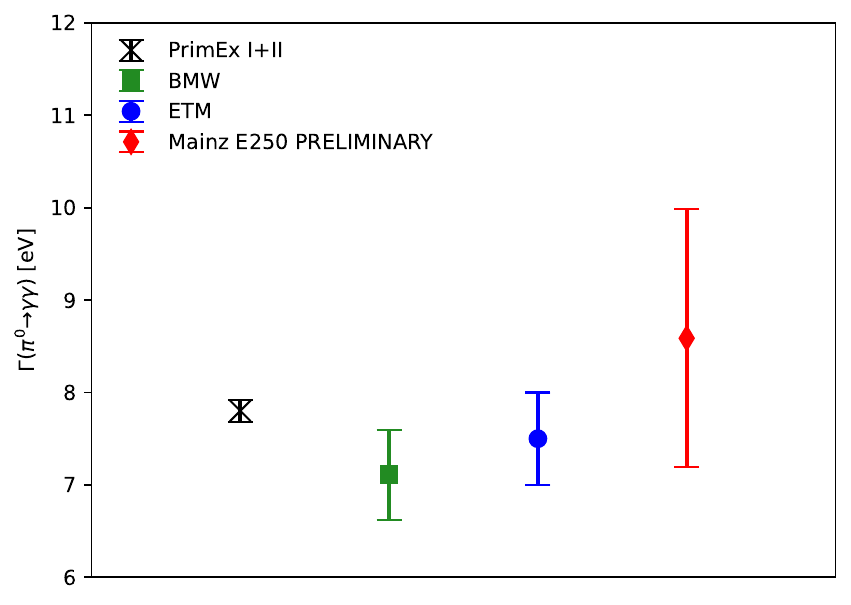}
\caption{Connecting the behaviour of the TFF at $Q^2=0$ with the partial decay width $\Gamma(\pi^0\to \gamma\gamma)$.
  The experimental result by PrimEx was published in~\cite{PrimEx-II:2020jwd}. We also compare to other lattice QCD
  results from the BMW collaboration~\cite{Gerardin:2023naa} and from the ETM collaboration~\cite{Alexandrou:2023lia}.}
\label{fig:smallQ2}
\end{figure}

Let us then recall the relation between the partial decay width $\Gamma(\pi^0\to \gamma\gamma)$
and the transition form factor:
\begin{equation}
  \Gamma(\pi^0\to \gamma\gamma) = \frac{\pi\alpha^2_e(m_{\pi^0})^3}{4}\mathcal{F}^2_{\pi^0\gamma^\ast\gamma^\ast}(0,0).
\label{eq:TFFandPartialDecayWidth}  
\end{equation}
This can be used to convert the normalization of the pion transition form factor into an estimate of the
decay width $\Gamma(\pi^0\to \gamma\gamma)$, and the results are depicted in Fig.~\ref{fig:smallQ2}.
The fit curves and data points plotted here are the same as in Fig.~\ref{fig:TFF} (but without the factor of
$Q^2$), and we simply zoom in on the small-$Q^2$ region. Comparison to the experimental result by PrimEx and
comparison with other lattice QCD results by BMW~\cite{Gerardin:2023naa} and ETMC~\cite{Alexandrou:2023lia}
shows we agree very well. Unfortunately, the lattice QCD results do not match the precision of the experimental
result yet.

\section{Summary and outlook}

In this report, we have given a status update of the Mainz group's calculation of the pion transition form
factor. The main development is the inclusion of a physical pion mass ensemble E250 with large volume.
We compute the disconnected diagrams needed in addition to the quark-line connected piece to construct
the full form factor. The new results presented here are still preliminary, and we plan to increase statistics.
At the final stage of the analysis, the result on E250 will be combined with the previous work published in
2019~\cite{Gerardin:2019vio} to extrapolate the form factor to the continuum and to physical quark masses.

The transition form factor $\mathcal{F}_{\pi^0\gamma^\ast \gamma^\ast}$ is the main ingredient in the estimation of the
pion-pole contribution to hadronic light-by-light scattering in the muon $g - 2$. The goal of this work is to
improve this estimate by including E250 in the analysis. However, if lattice QCD calculations want to address
the tension between the partial decay width $\Gamma(\pi^0\to \gamma\gamma)$ from the PrimEx-II experiment and
NLO theory predictions, there is still a long way to go.

\acknowledgments{
The authors acknowledge the support of Deutsche Forschungsgemeinschaft (DFG) through project
HI 2048/1-3 ``Precise treatment of quark-sea contributions in lattice QCD'' (project 399400745),
through the research unit FOR~5327 ``Photon-photon interactions in the Standard Model and beyond
-- exploiting the discovery potential from MESA to the LHC'' (grant 458854507),
and through the Cluster of Excellence ``Precision Physics, Fundamental Interactions and Structure of Matter''
(PRISMA+ EXC 2118/1) funded within the German Excellence Strategy (project ID 39083149).
We thank our colleagues in the CLS initiative for sharing ensembles.

We acknowledge PRACE for awarding us access to HAWK at GCS@HLRS, Germany via application 2020225457.
Parts of this research were conducted using the supercomputer MOGON 2 offered by Johannes Gutenberg
University Mainz (hpc.uni-mainz.de), which is a member of the AHRP (Alliance for High Performance
Computing in Rhineland Palatinate,  www.ahrp.info) and the Gauss Alliance e.V.
The authors gratefully acknowledge the Gauss Centre for Supercomputing e.V. (www.gauss-centre.eu) for
funding this project by providing computing time on the GCS Supercomputer SuperMUC-NG at Leibniz
Supercomputing Centre (www.lrz.de) and on the GCS Supercomputer JUWELS at Jülich Supercomputing Centre (JSC).
}

\end{document}